\documentclass[letterpaper,english,aps,prd,final,nofootinbib,twocolumn]{revtex4-1}
\usepackage[T1]{fontenc}
\usepackage[latin9]{inputenc}
\usepackage{babel}
\usepackage{amsmath}
\usepackage{amssymb}
\usepackage{graphicx}
\usepackage{wasysym}
\usepackage[unicode=true,
 bookmarks=false,
 breaklinks=false,pdfborder={0 0 1},backref=section,colorlinks=false]
 {hyperref}

\makeatletter

\pdfpageheight\paperheight
\pdfpagewidth\paperwidth

\usepackage[caption=false]{subfig}
\usepackage{tikz}

\makeatother

\begin{document}

\title{Cosmologically Viable Low-energy Supersymmetry Breaking}

\author{Anson Hook}
\email{hook@umd.edu}

\affiliation{Maryland Center for Fundamental Physics, Department of Physics\\
University of Maryland, College Park, Maryland 20742, USA}

\author{Robert McGehee}
\email{robertmcgehee@berkeley.edu}

\affiliation{Department of Physics, University of California, Berkeley, California
94720, USA}

\affiliation{Theoretical Physics Group, Lawrence Berkeley National Laboratory,
Berkeley, California 94720, USA }

\author{Hitoshi Murayama}
\email{hitoshi@berkeley.edu}

\email{hitoshi.murayama@ipmu.jp}

\affiliation{Department of Physics, University of California, Berkeley, California
94720, USA}

\affiliation{Theoretical Physics Group, Lawrence Berkeley National Laboratory,
Berkeley, California 94720, USA }

\affiliation{Kavli Institute for the Physics and Mathematics of the Universe (WPI),
University of Tokyo Institutes for Advanced Study, University of Tokyo,
Kashiwa 277-8583, Japan}

\date{\today}
\begin{abstract}
A recent cosmological bound on the gravitino mass, $m_{3/2}<4.7$~eV,
together with LHC results on the Higgs mass and direct searches, excludes
minimal gauge mediation with high reheating temperatures. We discuss
a minimal, vector-mediated model which incorporates the seesaw mechanism
for neutrino masses, allows for thermal leptogenesis, ameliorates
the $\mu$ problem, and achieves the observed Higgs mass and a gravitino
as light as 1\textendash 2~eV. 
\end{abstract}
\maketitle

\section{Introduction}

Supersymmetry (SUSY) is a well motivated theory beyond the standard
model. It offers a solution to the hierarchy problem and allows for
gauge unification (see, \textit{e.g.}\/, \cite{Murayama:2000dw}).
Models with SUSY breaking at low energies are especially interesting
because they imply light gravitinos which can be produced at colliders,
enabling experimental tests of the SUSY-breaking mechanism \cite{Cabibbo:1981er}.
Gauge mediation is the most studied way to do this \cite{AlvarezGaume:1981wy,Dine:1981gu,Nappi:1982hm}.

However, the window on usual gauge mediation is quickly closing as
cosmological and LHC bounds eliminate parameter space from both ends.
Decreasing upper bounds on the gravitino mass from cosmological data
are decreasing the upper bound on the SUSY-breaking scale, $\sqrt{F}$,
because $m_{3/2}=F/\sqrt{3}M_{Pl}$.
Increasing bounds on the gaugino masses from the LHC and the 125~GeV
Higgs mass are simultaneously increasing the lower bound on $\sqrt{F}$
. 

For most of the gravitino-mass range, if it is the stable Lightest
Supersymmetric Particle (LSP), it overcloses the Universe without
an unnaturally low reheating temperature \cite{Moroi:1993mb,deGouvea:1997afu}.
Such a low reheating temperature makes baryogenesis difficult as well.
In particular, thermal leptogenesis requires $T_{R}>10^{9}$ GeV \cite{Buchmuller:2004tu}.
A very light gravitino, $m_{3/2}<0.24$~keV, does not overclose the
Universe even when it is thermalized. But it constitutes a hot (or
warm) dark matter component and suppresses the structure of the Universe
at small scales. According to a very recent study, CMB lensing and
cosmic shear constrain its mass to be $m_{3/2}<4.7\,\text{eV}$ \cite{Osato}.
\begin{figure*}[!htp]
\includegraphics[width=\textwidth]{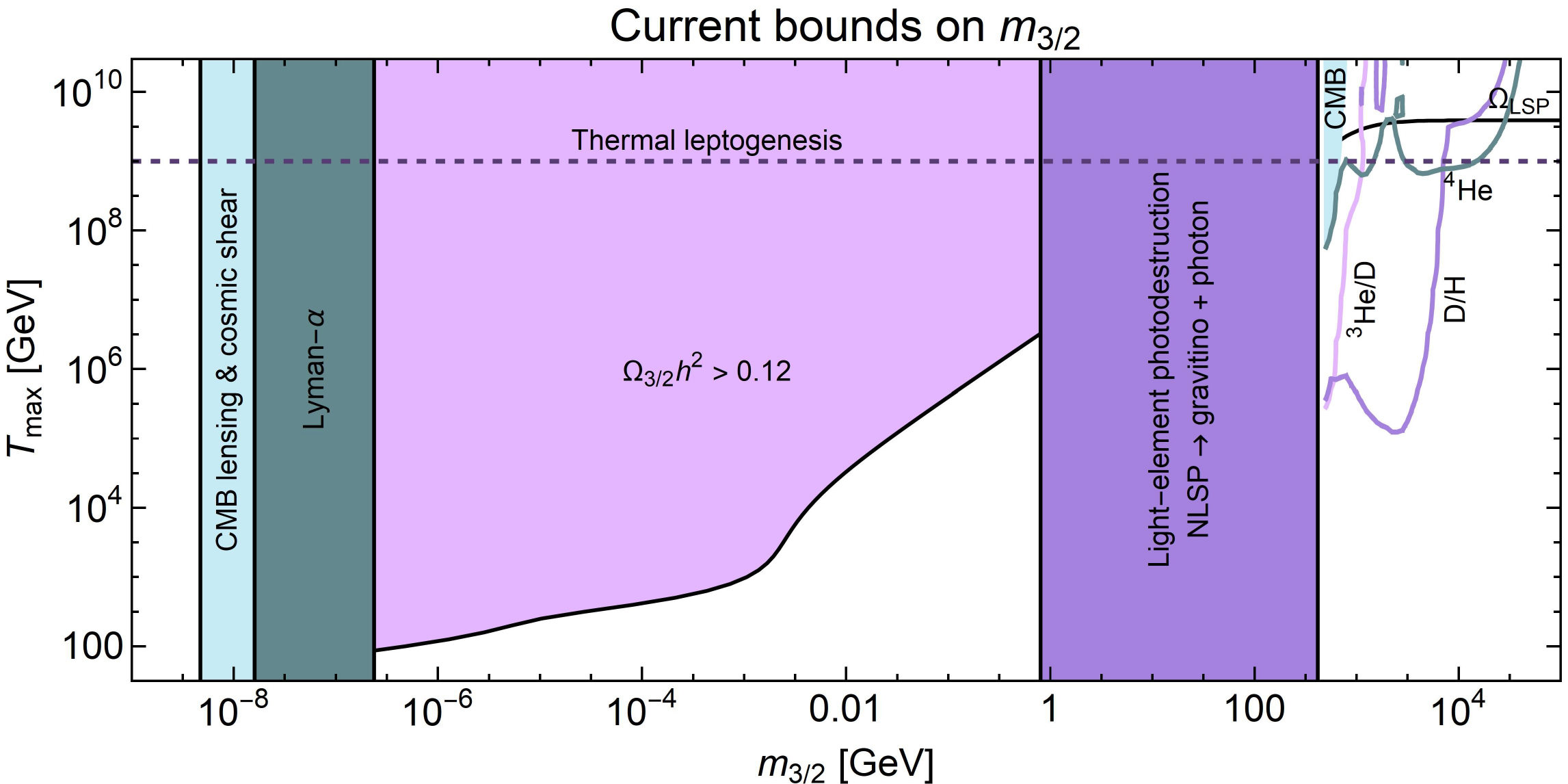}

\caption{\label{fig:introfig}A compilation of bounds in the $m_{3/2}$-$T_{max}$
plane, where $T_{max}$ is the maximum temperature from which the
usual, radiation-dominated universe starts. In inflationary models,
$T_{max}$ corresponds to the reheat temperature, $T_{R}$. The bound
from CMB lensing and cosmic shear is in light blue \cite{Osato}.
The Lyman-$\alpha$ bound from Ref.~\cite{Viel:2005qj} is in dark
green. The overclosure bound for gravitinos heavier than $\apprge0.2$
keV is in light purple \cite{Moroi:1993mb,Viel:2005qj}. We recalculated
this limit with current measurements of the Hubble scale factor and
dark matter abundance assuming $M_{1}=417\,\text{GeV}$ and the ratio
$M_{1}:M_{2}:M_{3}=1:2:7$. The bound from light-element photodestruction
is in dark purple \cite{Kawasaki:2008qe}. The bounds from a long-lived
gravitino affecting light-element abundances after BBN are taken from
Fig.~15 of Ref.~\cite{Kawasaki:2017bqm} for $m_{LSP}=M_{1}=417\,\text{GeV}$.
The lower bound on the reheat temperature from thermal leptogenesis
is the dark-purple, dashed line \cite{Buchmuller:2004tu}.}
\end{figure*}
Fig.~\ref{fig:introfig} summarizes these cosmological bounds on
$m_{3/2}$, as well as others, and demonstrates the shrinking parameter
space for any models with gravitinos lighter than roughly $10^{5}$
GeV.

In minimal, gauge-mediated models, on the other hand, the 125 GeV
Higgs mass requires a large stop mass. This implies $\sqrt{F}\gtrsim10^{3}\,\text{TeV}$.
Hence, the gravitino is heavier than 360~eV for one messenger or
60~eV for five messengers, which is the maximum number allowed by
perturbative gauge unification \cite{Ajaib:2012vc}. Therefore, minimal,
gauge-mediated models are excluded if the reheating temperature is
high.

This gravitino problem is often ignored in literature which attempts
to achieve the 125 GeV Higgs mass via gauge mediation, \textit{e.g.}\/,
using $A$ terms from RGE flow \cite{Draper:2011aa} or non-decoupling
$D$ terms \cite{Batra:2003nj,Maloney:2004rc}. A non-minimal Higgs
sector or strongly coupled messengers may achieve $m_{3/2}\lesssim16$~eV
\cite{Yanagida:2012ef}, but it is unclear if the gravitino mass can
be pushed below 4.7~eV. These problems can be ameliorated by using
the Next-to-Minimal Supersymmetric Standard Model (NMSSM) or Dirac-NMSSM \cite{Lu:2013cta}, coupling the Higgs to
messengers \cite{Craig:2012xp}, or having a strongly coupled messenger
sector \cite{Yanagida:2010zz,Heckman:2011bb,Evans:2012uf,Kitano:2012wv}.
However, these models are quite intricate.

In this paper, we elaborate on a simple, vector-mediated model proposed
by Hook and Murayama which breaks SUSY at tree level with a $U$(1)
$D$ term and evades all gravitino bounds \cite{Hook}. We discuss
electroweak symmetry breaking (EWSB) in detail to obtain the correct
Higgs mass while avoiding all LHC bounds on Minimal Supersymmetric Standard Model (MSSM) particles. We also
introduce a new, $U$(1) charge assignment where the right-handed
neutrino is neutral to allow the seesaw mechanism for neutrino masses
and thermal leptogenesis \cite{Davidson:2008bu}. Additionally, this
model produces a split spectrum usually found in gravity-mediated
models \cite{Evans:2015bxa}.

We first review the basics of SUSY breaking via vector mediation.\footnote{Not to be confused with the identically named theory in Ref.~\cite{Brummer:2006dg}
which has a different mechanism.} If we want a cosmologically viable, light gravitino and a naturally
high reheating temperature, Fig.~\ref{fig:introfig} clearly shows
it must be \textit{very}\/ light.\footnote{An alternative way to achieve a high reheating temperature is anomaly
mediation \cite{Randall:1998uk,Giudice:1998xp}.} A first attempt is to use a low-energy, gauge-mediated model. However,
normal, gauge-mediated models have gaugino and scalar masses at loop
level. Then the low, SUSY-breaking scale results in spartners too
light for the current LHC bounds. Thus, we want a low-energy, gauge-mediated
model with scalar masses at tree level.

However, there is a ``no-go theorem'' against models which break
SUSY at tree level \cite{Girardello:1981wz,Ferrara:1979wa,lawrence}.
The problem is the tree-level identity $\text{STr}\left(\mathcal{M}^{2}\right)=0$
which usually implies SUSY breaking cannot occur at the tree level.
This is because anomaly cancellation requires both positive and negative
$U$(1) charges which gives some scalars negative soft masses at tree
level from a $U$(1) $D$ term.

In vector mediation, there are vector-like messenger fields.\footnote{Refs. \cite{Nardecchia:2009ew,Monaco:2011fe} also used a $U$(1)
$D$ term to mediate SUSY breaking at tree level, but at the GUT scale.
This resulted in a very weakly coupled gravitino and thus, very different
collider and cosmology signatures \cite{Arcadi:2011yw}. } We assign positive $U$(1) charges to all sfermions and negative
charges to all vectorlike messengers. Their vectorlike masses overcome
their negative soft masses so that no scalars are tachyonic. Thus,
vector mediation allows tree-level scalar masses and a lower SUSY-breaking
scale. This gives a lighter gravitino while getting the Higgs mass
correct and avoiding current bounds on MSSM particles. The lower scale
also ameliorates the so-called $\mu$ problem.

\section{The Models}

The vector-mediated models \cite{Hook} employ an $E_{6}$-inspired
particle content that consists of three families of the fundamental
representation decomposed into $SO(10)\times U(1)_{\psi}$ as 
\begin{equation}
\boldsymbol{\boldsymbol{27}}=\Psi(\boldsymbol{16},+1)\oplus\Phi(\boldsymbol{10},-2)\oplus S(\boldsymbol{1},+4),\label{eq:E6decomp}
\end{equation}
where $\Psi$ contains the SM fermions and $\Phi$ contains two families
of messengers and the MSSM Higgs doublets with their color triplets.
Because the messengers and Higgs color triplets have different couplings
than the Higgs doublets, we distinguish between generations using
subscripts with $\Phi_{1}\equiv(T_{u},H_{u})+(T_{d},H_{d})$ where
$T$ are the color triplets. Since these Higgs color triplets' masses
are not at the grand-unified-theory scale, they prevent gauge unification.
However, we can add two new electroweak doublets, uncharged under
our $U$(1), with the same mass as the Higgs color triplets in order
to form complete $SU$(5) multiplets and restore gauge unification.
We also include a neutral particle $Z\left(0\right)$ and a vectorlike
multiplet charged under $U(1)_{\psi}$, $X\left(-4\right)$ and $Y\left(+4\right)$.
These particles are responsible for SUSY breaking. 

The superpotential for our model is
\begin{align}
W= & \,MSX+\lambda Z\left(XY-v_{S}^{2}\right) \nonumber \\
 & -2T_{u}{T}_{d}\left(\frac{g}{2}Y+\frac{k}{2}S\right)-2H_{u}H_{d}\left(\frac{g_{H}}{2}Y+\frac{k_{H}}{2}S\right)\nonumber \\
 & -\sum_{a=2}^{3}\Phi_{a}\cdot\Phi_{a}\left(\frac{g}{2}Y+\frac{k}{2}S\right)\nonumber \\
 & +\frac{\lambda^{S}}{2}\sum_{a=2}^{3}\Phi_{a}\cdot\Phi_{a}\sum_{b=1}^{2}S_{b}+\kappa X\sum_{a=1}^{2}N_{a}S_{a}.\label{eq:W} 
\end{align}
The three generations of $S$ are denoted $S_{1}$, $S_{2}$ and $S$.
The first line in Eq.~\eqref{eq:W} breaks SUSY with $F$ terms
and a positive $D$ term. 
It can be dynamically realized as the low-energy effective theory of an Izawa-Yanagida-Intriligator-
Thomas (IYIT) model with 4 $SU(2)$ doublets with appropriate $U(1)$ charges \citep{Izawa:1996pk,Intriligator:1996pu}.
The second and third lines are standard
messenger interactions and generate the usual, gauge-mediated gaugino
masses at one loop. We have allowed the MSSM Higgs doublets, $H_{u}$
and $H_{d}$, to have different couplings to $Y$ and $S$ than their
color triplets, $T_{u}$ and $T_{d}$, in order to satisfy EWSB conditions.
Each generation of $\Phi$ could have different couplings to $Y$
and $S$, but we set them equal to maximize the gaugino masses (see
\ref{sec:Constraints-prior-to}). The fourth line gives the $S_{1,2}$
fermions masses by introducing two neutral fields $N_{1,2}$, while
the $S_{1,2}$ scalars also acquire a mass from the $D$ term. We
have the interaction $W\supset S_{1,2}\Phi\Phi$ in line four to allow
$S_{1,2}$ and $N_{1,2}$ to decay so that they don't overclose the
universe.

In this paper, we consider two different charge assignments under
a new $U(1)$: the original-charge assignment corresponding to $U(1)_{\psi}$
and a new, seesaw-charge assignment corresponding to a different $U(1)_{SS}$.
For clarity, we distinguish between these charge assignments using
these subscripts and omit a subscript when talking about either charge
assignment generally. We explore the viability of models with each
of these charge assignments separately.

The right-handed neutrino has charge $+1$ under $U(1)_{\psi}$, as
do all SM fermions by virtue of Eq.~\eqref{eq:E6decomp}. Thus, gauging
$U(1)_{\psi}$ does not allow the seesaw mechanism to generate neutrino
masses. However, the right-handed neutrino can be made neutral by
instead gauging a linear combination of $U(1)_{\psi}$ and $U(1)_{\chi}$,
where a given $SO(10)$ representation is decomposed into $SU(5)\times U(1)_{\chi}$.
The linear combination we gauge is 
\begin{equation}
Q_{SS}=\frac{1}{4}\left(5Q_{\psi}+Q_{\chi}\right).\label{eq:U1charge}
\end{equation}
Using this charge, we find the decomposition of $\Psi(\boldsymbol{16},+1)$
into $SU(5)\times U(1)_{SS}$ is
\begin{equation}
\boldsymbol{16}=\boldsymbol{10}(+1)\oplus\boldsymbol{\bar{5}}(+2)\oplus\boldsymbol{1}(0).\label{eq:16decomp}
\end{equation}
The right-handed neutrino of each family of $\Psi$ is neutral under
$U(1)_{SS}$ by virtue of Eq.~\eqref{eq:16decomp}, allowing the seesaw
mechanism for neutrino masses. The decomposition of $\Phi(\boldsymbol{10},-2)$
into $SU(5)\times U(1)_{SS}$ is 
\begin{equation}
\boldsymbol{10}=\boldsymbol{5}(-2)\oplus\boldsymbol{\bar{5}}(-3).\label{eq:10decomp}
\end{equation}
Additionally, each $S$ has charge $Q_{SS}=5$, $X(Q_{SS}=-5)$, $Y(Q_{SS}=+5)$,
and $Z$ remains neutral.

We do not have the usual $\mu$-term as it is not gauge invariant
under $U(1)_{\psi}$ or $U(1)_{SS}$. Similar to the NMSSM, it is generated by the
expectation values of $Y$ and $S$. For the two charge assignments, we have to 
introduce different interactions and discrete symmetries to allow the color triplets
to decay while preventing proton decay. 

For the seesaw-charge assignment, we introduce the interaction
$W\supset \frac{1}{M_{Pl}}SQ T_{d} H_{d}$ to allow for the color triplets in
$\Phi$ to decay. We assign negative
matter-parity to all SM multiplets in $\Psi$ and to the color triplets
in $\Phi$, but positive matter-parity to the electroweak doublets
in $\Phi$ to prevent proton decay. \footnote{R. M. thanks T. Yanagida for pointing out this issue. Matter-parity assignment is related to R-parity
assignment via $P_{M}=P_{R}\times (-1)^{2s},$ where $s$ is the spin of the particle in question.} This is
consistent with assigning $B=+1/3$ to $T_u$ and $B=-1/3$ to $T_d$.

For the original-charge assignment, we introduce the interaction
$W\supset T_{d} \bar{u} \bar{d}$ to allow for the color triplets in
$\Phi$ to decay. We enforce lepton-number conservation to prevent proton decay.
Since the right-handed neutrinos are charged under $U(1)_{\psi}$, this requirement
is consistent with the neutrinos having Dirac masses.

For the seesaw-charge assignment in Eq.~\eqref{eq:10decomp}, the
MSSM Higgs soft masses satisfy $m_{H_{u}}^{2}>m_{H_{d}}^{2}$ at tree
level since the $D$ term is positive and $-2>-3$. This is problematic
for EWSB. To help, we introduce kinetic mixing between our $U(1)_{SS}$
(or $U(1)_{\psi}$ for the original-charge assignment) and $U(1)_{Y}$.
Following Ref.~\citep{Dienes}, this kinetic mixing can be written
as a cross term between the $D$ terms associated with the two $U(1)$'s
as 
\begin{equation}
V\supset g'\chi e\sum_{i,j}Q_{Y}^{i}Q^{j}\left|\phi_{i}\right|^{2}\left|\phi_{j}\right|^{2},\label{eq:kinmixing}
\end{equation}
where $Q_{Y}^{i}$ is the $U(1)_{Y}$ charge of the scalar $\phi_{i}$,
$Q^{j}$ is the $U(1)$ charge, and $e$ is the $U(1)$ electric charge.
The dimensionless coupling $\chi$ is a measure of the mixing and
we have dropped terms proportional to $\chi^{2}$.
$\chi$ is naturally small and is generated by 1-loop diagrams with particles charged under both $U(1)$'s \citep{Dienes}.
Note that the addition
of this mixing does not change the locations of the false and true
vacuua. The $D$ term gives every scalar $\phi_{i}$ a mass contribution
from Eq.~\eqref{eq:kinmixing}, 
\begin{equation}
\delta m_{i}^{2}=Q_{Y}^{i}g'\chi\left\langle D\right\rangle .\label{eq:delm}
\end{equation}
Since $H_{u}$ and $H_{d}$ have opposite $U(1)_{Y}$ charges, Eq.~\eqref{eq:delm} aids EWSB in models with the seesaw-charge assignment.

\section{\label{sec:Constraints-prior-to}Constraints prior to EWSB}

We explore the parameter space $(e,M,\lambda,v_{S},\chi,k,g,k_{H},g_{H},\tan\beta)$
for both charge assignments to find viable models with light gravitinos
which avoid the cosmological bound $m_{3/2}<4.7\,\text{eV}$. Before
we verify EWSB and the Higgs mass, we must impose some constraints
on our parameter space. The first suppresses tunneling to the true
vacuum.

The interactions in lines two and three of Eq.~\eqref{eq:W} which
give rise to the gaugino masses at one loop also generate a supersymmetric
vacuum. This true vacuum appears at 
\begin{equation}
\begin{array}{ccc}
X_{0}=\dfrac{v_{s}^{2}}{Y_{0}}=\dfrac{k\Phi_{0}^{2}}{2M}, & Z_{0}=\dfrac{gM}{k\lambda}, & S_{0}=-\dfrac{2gMv_{S}^{2}}{k^{2}\Phi_{0}^{2}}\end{array},
\end{equation}
where $\Phi_{0}$ is set by requiring the $D$ term vanishes. We require
the parameter space we consider to disallow substantial vacuum or
thermal tunneling between the false and true vacuua.

To calculate the vacuum tunneling rate, we calculate the bounce profile
$\bar{\phi}(r)$ which solves 
\begin{equation}
\frac{d^{2}\bar{\phi}_{i}}{dr^{2}}+\frac{3}{r}\frac{d\bar{\phi}_{i}}{dr}=\frac{dV}{d\bar{\phi}_{i}}\label{eq:bouncediffeq}
\end{equation}
with the boundary conditions 
\begin{equation}
\begin{array}{cc}
\dfrac{d\bar{\phi}_{i}}{dr}\left(r=0\right)=0, & \bar{\phi}_{i}\left(r\rightarrow\infty\right)=\phi_{0}\end{array},\label{eq:bounceBCs}
\end{equation}
where $\phi_{0}=\left(X_{0},Y_{0},0,0,0\right)$ is the false vacuum.
The spherically symmetric, 4D Euclidean action of a profile $\phi\left(r\right)$
is 
\begin{equation}
S_{E}\left(\phi\left(r\right)\right)=2\pi^{2}\int_{0}^{\infty}drr^{3}\left[\sum_{i}\frac{1}{2}\left(\frac{d\phi_{i}}{dr}\right)^{2}+V\left(\phi\left(r\right)\right)\right],\label{eq:S_E}
\end{equation}
where the sum is over all field dimensions $\left(X,Y,Z,S,\Phi\right)$.
The tunneling rate per unit volume is exponentially sensitive to the
bounce action $B=S_{E}\left(\bar{\phi}\right)-S_{E}\left(\phi_{0}\right)$
as $\Gamma\propto\exp\left(-B\right)$ \citep{Coleman}.

The thermal bounce action, $B_{th}=S_{th}\left(\bar{\phi}\right)-S_{th}\left(\phi_{0}\right)$,
is similar: the action is given by the 3D version of \eqref{eq:S_E}
and the bounce profile solves the 3D versions of \eqref{eq:bouncediffeq}
and \eqref{eq:bounceBCs}. The thermal tunneling rate is $\Gamma_{th}\propto T^{4}\exp\left(-B_{th}/T_{C}\right)$,
where $T_{C}$ is the critical temperature at which the SUSY-invariant
vacuum becomes the true minimum.

When approximating both the vacuum and thermal bounce actions, we
calculate the tunneling along the straight line between the true and
false vacuua. The 4D vacuum and 3D thermal bounce actions are calculated
semi-analytically using results from Ref.~\citep{Adams}. In order
for these tunneling rates to be sufficiently small, we conservatively
require that $B>450$ and $B_{th}/T_{c}>250$. We use the high-temperature
approximation for the thermal effective potential to calculate $T_{C}$.
We also assume $B_{th}$ is approximately constant and calculate it
using the zero-temperature potential. For much of our parameter space,
$T_{C}\apprle v_{S}$ where this approximation breaks down. However,
we find that $B_{th}$ is so large in our region of interest that
this does not matter. We verify that $\text{STr}\left(\mathcal{M}^{2}\right)=0$
along the straight line between the true and false vacuua. We set
$\kappa=0.1$ for concreteness.

To illustrate that the vacuum and thermal tunneling rates are highly
suppressed, we have chosen the representative values $\left(M=2,\lambda=4\pi,\chi=0,k=4\pi\right)$
and set $g$ as small as possible before the lightest messengers go
tachyonic (see Eq.~\eqref{eq:messmatr} below). See Fig.~\ref{fig:B}
and Fig.~\ref{fig:Bth}. 
\begin{figure}[!htp]
\includegraphics[width=\columnwidth]{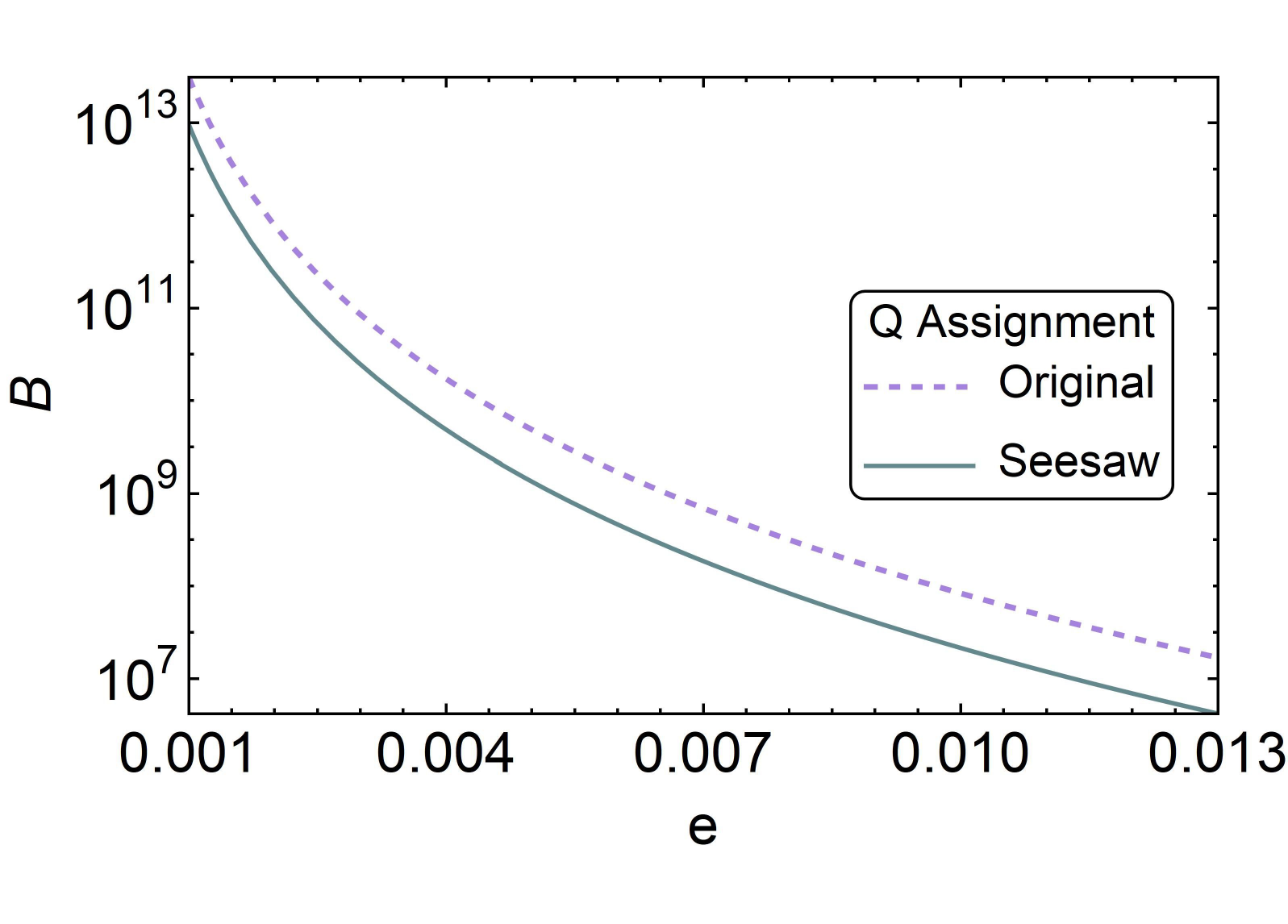}

\caption{\label{fig:B}Vacuum bounce action for both charge assignments. We
take the representative values $\left(M=2,\lambda=4\pi,\chi=0,k=4\pi\right)$
and set $g$ as small as possible before the lightest messengers go
tachyonic. We easily satisfy $B>450$ for the Universe to be stable.}
\end{figure}
\begin{figure}[!htp]
\includegraphics[width=\columnwidth]{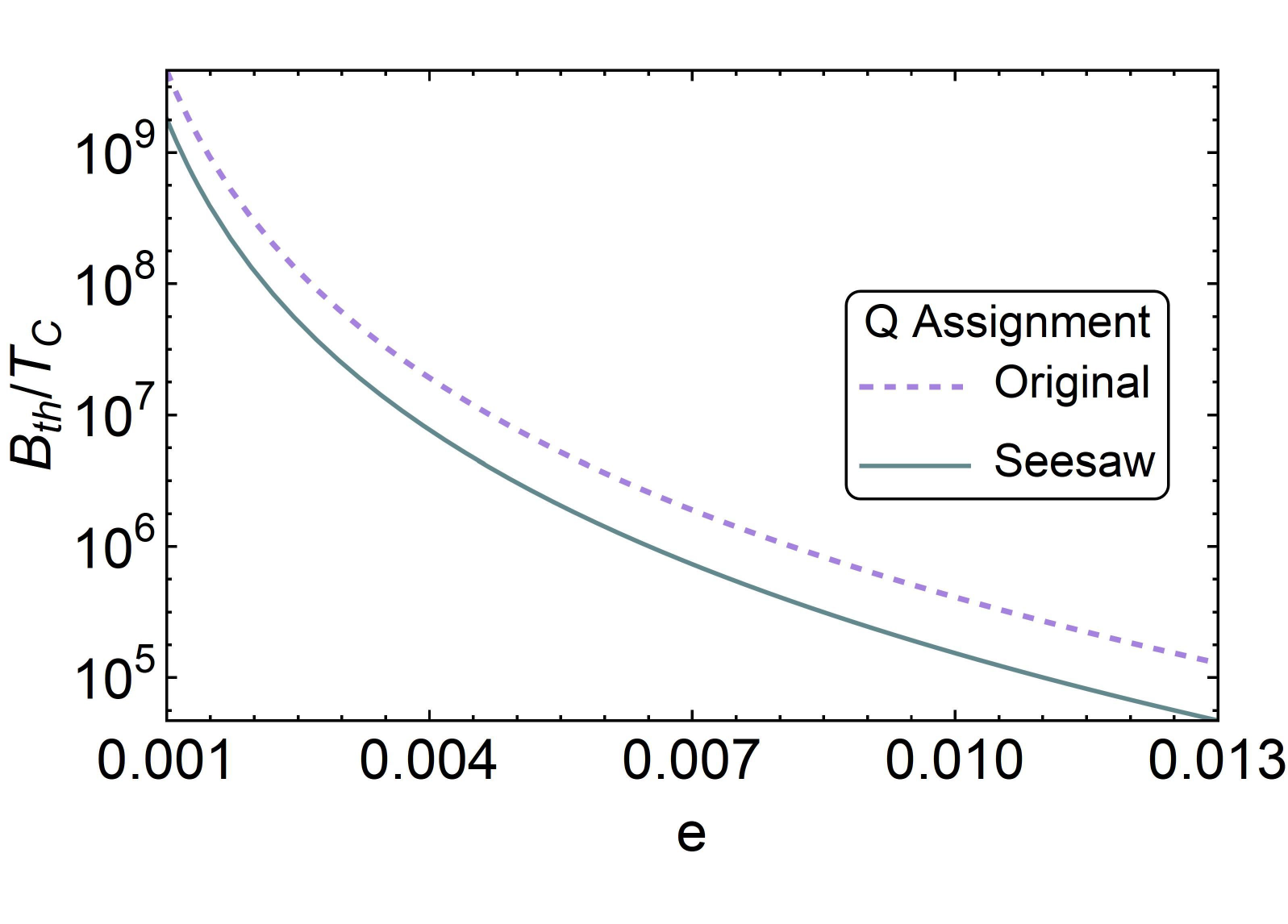}

\caption{\label{fig:Bth}$B_{th}/T_{C}$ for both charge assignments. We take
the same values as in Fig.~\eqref{fig:B}. We easily satisfy $B_{th}/T_{C}>250$
for the Universe to be stable.}
\end{figure}

Both vacuum and thermal bounce actions are approximately independent
of $\chi$ over the range of values we consider later. While $\chi=0$
is not viable for the seesaw-charge assignment due to EWSB failure,
these figures do not change for realistic values of $\chi$. We allow
different values of $\left(M,\lambda,\chi\right)$ when conducting
a full parameter-space search. However, we always take $k=4\pi$ and
minimize $g$ in order to maximize the gaugino masses (see \ref{sec:Constraints-prior-to}).

Both actions are dramatically larger than in Ref.~\citep{Hook} because
we are considering much smaller values of $e$. One can guess that
a smaller value of $e$ makes the bounce actions much larger because
when $e\rightarrow\infty$, $\Phi$ is tachyonic and there is no barrier
between the false and true vacuua. As there is a barrier for small
$e$, by continuity, one expects that smaller $e$ gives larger barriers.
In the limit $e\rightarrow0$, we numerically find that the barrier
height between the two vacuua increases. This is the thin-wall limit
and the bounce actions increase dramatically~\citep{Coleman,coleman1985,Kobzarev}.

The scalar potential along the straight-line, tunneling path is quartic
in one scalar field. This potential is characterized by a single,
dimensionless parameter $\delta$ which can take values between 0
and 2 \citep{Adams}. The thin-wall limit corresponds to the limit
$\delta\rightarrow2$. In this limit, the analytic, vacuum bounce
action is proportional to $\left(2-\delta\right)^{-3}$ and the analytic,
thermal bounce action is proportional to $\left(2-\delta\right)^{-2}$.
For the parameters taken in Ref.~\citep{Hook}, $\delta=1.6$. We
consider points in parameter space which are much closer to the thin-wall
limit with $\delta\apprge1.996$.

The gaugino masses are generated from one-loop diagrams with the messengers
in the three families of $\Phi$. All fermions in $\Phi$ have the
same mass $M_{\Phi}=gY_{0}$. Taking into account the kinetic mixing
in Eq.~\eqref{eq:delm}, the boson components all have the mass matrix
\begin{equation}
\setlength\arraycolsep{1pt}
\medmuskip = 1mu
\begin{pmatrix}
M_{\Phi}^{2}+Q_{H_{u}}eD+Q_{H_{u}}^{Y}g'\chi D & kF_{S}\\
kF_{S} & M_{\Phi}^{2}+Q_{H_{d}}eD-Q_{H_{u}}^{Y}g'\chi D
\end{pmatrix},\label{eq:messmatr}
\end{equation}
where $Q_{H_{u}}$ is the $U(1)$ charge of fields in the \textbf{5
}of $\Phi$, $Q_{H_{d}}$ is the charge of fields in the $\boldsymbol{\bar{5}}$
of $\Phi$, $F_{S}=MX_{0}$, and $D=eQ_{X}\left(X_{0}^{2}-Y_{0}^{2}\right)>0$.
$Q_{X}=-Q_{Y}$ since $X$ and $Y$ form a vectorlike multiplet. For
the bosons in the color triplets, this is the correct mass matrix
under the appropriate replacement $Q_{H_{u}}^{Y}=+\frac{1}{2}\rightarrow Q_{d}^{Y}=-\frac{1}{3}$.
We leave the $U(1)$ charges unspecified since we find viable models
for both charge assignments. The MSSM Higgses have the same mass matrix
under the replacement $k\rightarrow k_{H},g\rightarrow g_{H}$ (see
Eq.~\eqref{eq:W}).

We calculate the one-loop gaugino masses using Eq.~(2.3) in Ref.~\citep{Poppitz}.
All three families' color triplets and electroweak doublets contribute
to the gaugino masses. The on-shell gluino mass enhancement from (s)top
loops is calculated using SOFTSUSY (discussed below) \cite{Allanach,Allanach:2014nba,Harlander:2017kuc,Kant:2010tf}.

The gaugino masses are maximized when there is a large mass hierarchy
in Eq.~\eqref{eq:messmatr}. We ensure a large hierarchy by choosing
parameters where the lighter scalar is light ($\approx$ 1 TeV). The
gaugino masses are also proportional to the fermion masses so we prefer
larger $g$ and larger $Y_{0}$. In order to increase the gaugino
masses in units of $v_{S}$, we thus set $k=4\pi$ and choose $g$
such that the lighter scalar in Eq.~\eqref{eq:messmatr} is near 1
TeV. These are the prescriptions we used in Figs. \eqref{fig:B} and
\eqref{fig:Bth} and allow smaller $v_{S}$.

We set $v_{S}$ by satisfying all current, gaugino-mass bounds. The
ATLAS bound on the gluino mass is roughly 2.03 TeV \citep{ATLASgluino},
while the CMS bound is 1.95 TeV \citep{CMSgluino}. The ATLAS bound
on the Wino mass is roughly 620 GeV, unless the Bino is heavier than
350 GeV, in which case it is 720 GeV \citep{ATLASBino}. The Wino
bound is the most stringent from direct searches. So, we use it to
estimate $v_{S}$ before EWSB. But we find getting the correct Higgs
mass is generally harder and therefore, determines $v_{S}$ after
EWSB (see below).

The gravitino mass is given by 
\begin{equation}
m_{3/2}^{2}=\frac{V}{3M_{Pl}^{2}},
\end{equation}
where $M_{Pl}=2.4\times10^{18}$ GeV. For a viable model, we require
$m_{3/2}<4.7$ eV \citep{Osato}.

We always set $\chi$ negative to make $m_{H_{u}}^{2}$ lighter and
help EWSB (see Eq.~\eqref{eq:messmatr}). Since $\tilde{\bar{e}}$
has the greatest $U(1)_{Y}$ charge, $Q_{\bar{e}}=+1$, we determine
how negative $\chi$ can be by requiring that $\tilde{\bar{e}}$ is
not tachyonic.

We also require that the massive $U(1)$ boson does not mix too much
with $W$ and $Z$ to affect the $\rho$ parameter.\footnote{R. M. thanks Simon Knapen on this point.}
We conservatively require that $\delta\rho<10^{-3}$ which imposes
that $\left(\frac{ev}{M_{V}}\right)^{2}<10^{-3}$, where $v$ is the
Higgs vacuum expectation value and $M_{V}$ is the mass of our $U(1)$
boson \cite{Patrignani:2016xqp}.

\section{Constraints after EWSB}

We use SOFTSUSY to calculate radiative EWSB.
Our inputs into SOFTSUSY are the gaugino masses, trilinear couplings,
sfermion masses, $m_{H_{u}}^{2}$, $m_{H_{d}}^{2}$, and $\tan\beta$.
We can read off the MSSM Higgs parameters from Eq.~\eqref{eq:messmatr}:
\begin{equation}
\begin{array}{ccc}
\mu=g_{H}Y_{0}, &  & m_{H_{u}}^{2}=Q_{H_{u}}eD+\frac{1}{2}g'\chi D,\\
B\mu=k_{H}F_{S}, &  & m_{H_{d}}^{2}=Q_{H_{d}}eD-\frac{1}{2}g'\chi D.
\end{array}
\end{equation}
The soft, SUSY-breaking $A$ terms are all $0$ in our models at tree
level. SOFTSUSY effectively sets $\mu$ and $B\mu$ as output boundary
conditions and therefore, $k_{H}$ and $g_{H}$ in our models.

We only input the above boundary conditions at tree level. We input
the gaugino masses at one-loop level, but do not include QCD enhancements
to $M_{3}$ or other, higher-order corrections. This is slightly inconsistent
as we run SOFTSUSY with the 3-loop renormalization group equations
and calculate the lightest MSSM Higgs both at 2-loop and 3-loop level.
When calculating the Higgs at 3-loop, SOFTSUSY also defaults to adding
2-loop Yukawa and $g_{3}$ threshold corrections to $m_{t}$ and $m_{b}$.

EWSB itself is a strong constraint. If a point in parameter space
has successful radiative EWSB, there are still many additional constraints
we check before considering it viable. In order to approximate EWSB
using SOFTSUSY, the correction to the MSSM, Higgs-quartic coupling,
$\lambda_{h}=m_{h}^{2}/v^{2}$, must be sufficiently small. We require
our Higgs mass to be accurate to $0.5$ GeV. This places an upper
bound at roughly 
\begin{equation}
\delta\lambda_{h}=g_{H}^{2}+\frac{1}{2}Q_{H_{d}}^{2}e^{2}+\frac{1}{2}eg'\chi<0.002.\label{eq:dellam}
\end{equation}
This is a conservative limit because the quartic couplings due to
the $F$ term and the $D$ term of heavy fields decouple and the effect
is in general smaller. We do not bother including a term proportional
to $k_{H}^{2}$ since $k_{H}^{2}\ll g_{H}^{2}$ generically. Eq.~\eqref{eq:dellam}
is also conservative because the $g_{H}^{2}$ piece vanishes in the
limit of large or small $\tan\beta$. Due to the upper bound on $\delta\lambda_{h}$,
Eq.~\eqref{eq:dellam} lets us only consider $e<0.021$. We find viable
models only occur for smaller values of $e$. This a nontrivial check
because $(k_{H},g_{H})$ are set by the output boundary conditions
from SOFTSUSY.

Requiring the Higgs mass to be $125$ GeV is the strongest constraint
and sets the overall scale $v_{S}$ in our models. If $v_{S}$ is
set to satisfy the gaugino constraints, the stop masses are too light
to lift the Higgs mass. When searching for viable models, we estimate
$v_{S}$ by satisfying the Wino constraint and then explore parameter
space by increasing this $v_{S}$ by some factor between $\sim2-3$.

We require that the output top Yukawa is still perturbative. We calculate
$y_{t}$ using Eq.~(62) in Ref.~\citep{Degrassi} using the top mass
in Ref.~\citep{ATLAStop}. We require that $y_{t}<0.94$. We also
require that the other neutral scalar Higgses are not too light. As
a conservative constraint, we require the mass of the heavier CP-even
neutral scalar, $H^{0}$, to be larger than $580$ GeV \citep{Craig}.
We require the mass of the CP-odd neutral scalar, $A^{0}$, to avoid
the $\tan\beta$-dependent, ATLAS bounds in Fig.~10 b of Ref.~\cite{ATLASmAvstanb}.

Since SOFTSUSY includes loop corrections to the gluino, we make sure
the output gluino is still heavier than 2.03 TeV. SOFTSUSY also gives
the spectrum of the neutralinos and charginos. We make sure to avoid
the bound on $\tilde{\chi}_{1}^{\pm}$ and $\tilde{\chi}_{2}^{0}$
as a function of $\widetilde{\chi}_{1}^{0}$ given in Ref.~\cite{ATLASBino}.

\section{SUSY Spectra and Best Models}

\newlength\temph
\begin{figure*}[!htp]%
\settoheight{\temph}{\includegraphics[width=0.44\textwidth]{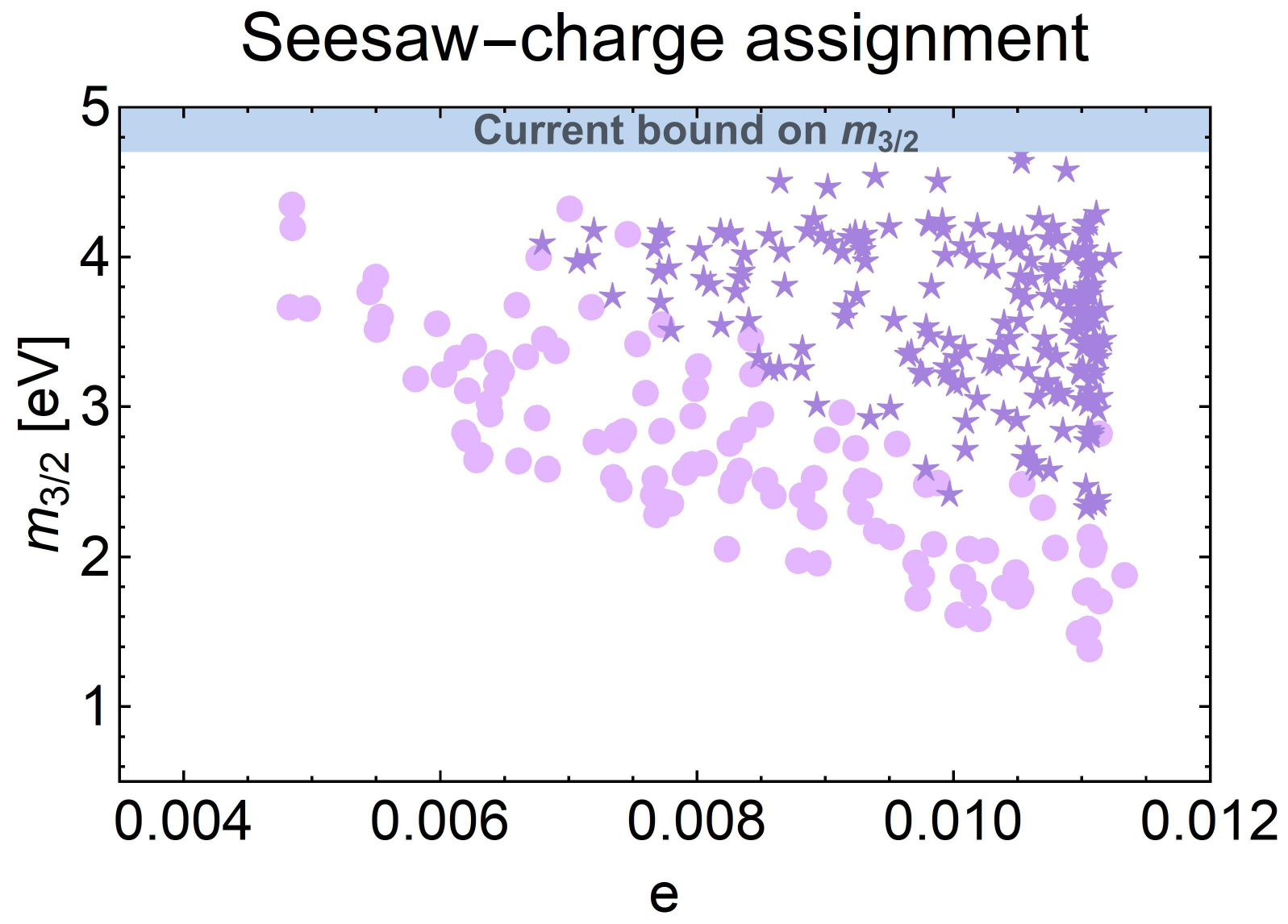}}
\centering
\subfloat{\includegraphics[width=0.44\textwidth]{SSQGoodfig.pdf}}%
\hfil
\subfloat{\includegraphics[width=0.44\textwidth]{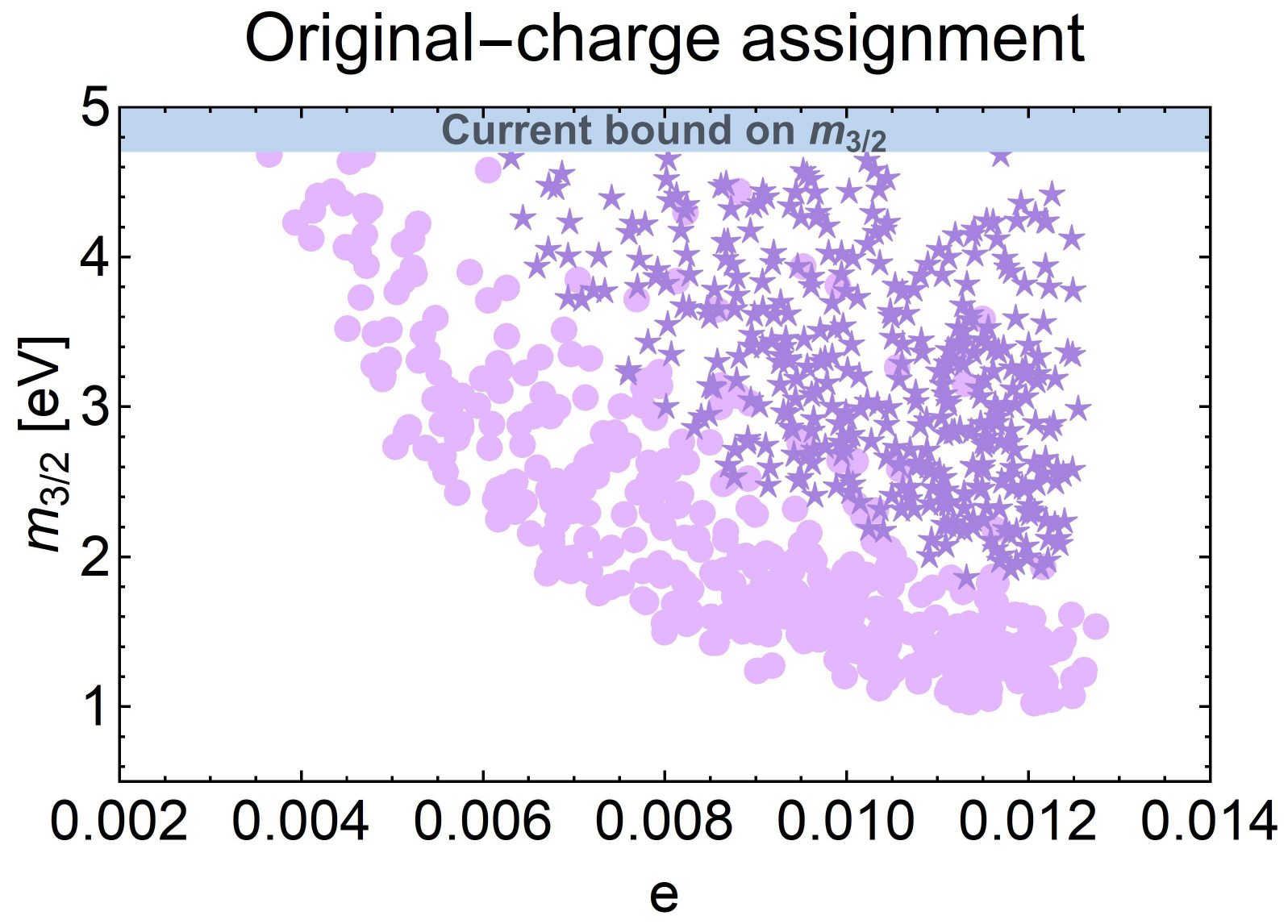}}%
\hfil
\subfloat{\tikz\node[minimum height=\temph]{\includegraphics[width=0.1\textwidth]{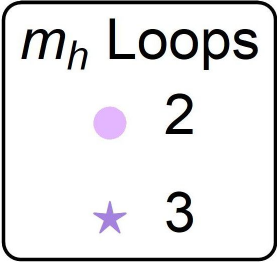}}; }%
\caption{$m_{3/2}$ vs. $e$ for viable models with the seesaw-charge and original-charge assignments
which satisfy all aforementioned constraints. Viable points are shown
when the Higgs is calculated at both the 2-loop and 3-loop level in
SOFTSUSY.}
\label{fig:goodfig}
\end{figure*}
Fig.~\ref{fig:goodfig} shows our search through parameter
space for viable models with the seesaw-charge assignment, $Q_{SS}$, and the original-charge assignment, $Q_{\psi}$.
The points correspond to viable points in the parameter space $(e,M,\lambda,v_{S},\chi,k,g,k_{H},g_{H},\tan\beta)$
for which all of the constraints are satisfied. We show viable points
when we calculate the Higgs mass at both the 2-loop and 3-loop level
in SOFTSUSY. The 3-loop calculation generally yields a lighter Higgs
which requires a greater $v_{S}$ to obtain $125$ GeV. This yields
a heavier gravitino and explains the separation of viable points between
the 2-loop and 3-loop calculations.
We find $v_S \in \left[55, 240 \right]$ TeV for the seesaw-charge assignment and $v_S \in \left[44, 240 \right]$ TeV for the original-charge assignment.

For the seesaw-charge assignment, $\chi$ is always close to its maximum
value to help alleviate $m_{H_{u}}^{2}>m_{H_{d}}^{2}$, namely $\chi \in \left[ -0.033,-0.014 \right]$.
This is naturally small and generated by one-loop diagrams of particles charged under both $U(1)$'s \cite{Dienes}.
However, it
is not possible to take $\chi$ large enough to make $m_{H_{u}}^{2}\leq m_{H_{d}}^{2}$
without making $\tilde{\bar{e}}$ tachyonic. EWSB works due to negative,
one-loop corrections to $m_{H_{u}}^{2}$ from (s)tops at the geometric
mean of the stop masses. In particular, the one-loop corrections to
$m_{H_{u}}^{2}$ from the (s)tops are 
\begin{align}
\Delta m_{H_{u}}^{2}= & \frac{N_{C}y_{t}^{2}}{\left(4\pi\right)^{2}}\Biggl(m_{\tilde{t}_{R}}^{2}\ln\left(\frac{m_{\tilde{t}_{R}}^{2}}{\mu_{R}^{2}}\right)+m_{\tilde{t}_{L}}^{2}\ln\left(\frac{m_{\tilde{t}_{L}}^{2}}{\mu_{R}^{2}}\right) \nonumber \\
 & -m_{\tilde{t}_{R}}^{2}-m_{\tilde{t}_{L}}^{2}\Biggr),\label{eq:mHu1loop}
\end{align}
where $N_{C}=3$ is the number of colors. At the renormalization scale,
$\mu_{R}^{2}=m_{\tilde{t}_{R}}m_{\tilde{t}_{L}}$, this correction
is negative as long as the stop masses are not too separated. This
negative correction, in addition to the kinetic mixing $\chi$, enables
EWSB in models with the seesaw-charge assignment.
\newlength\ttemph
\begin{figure*}[!htp]%
\settoheight{\ttemph}{\includegraphics[width=0.44\textwidth]{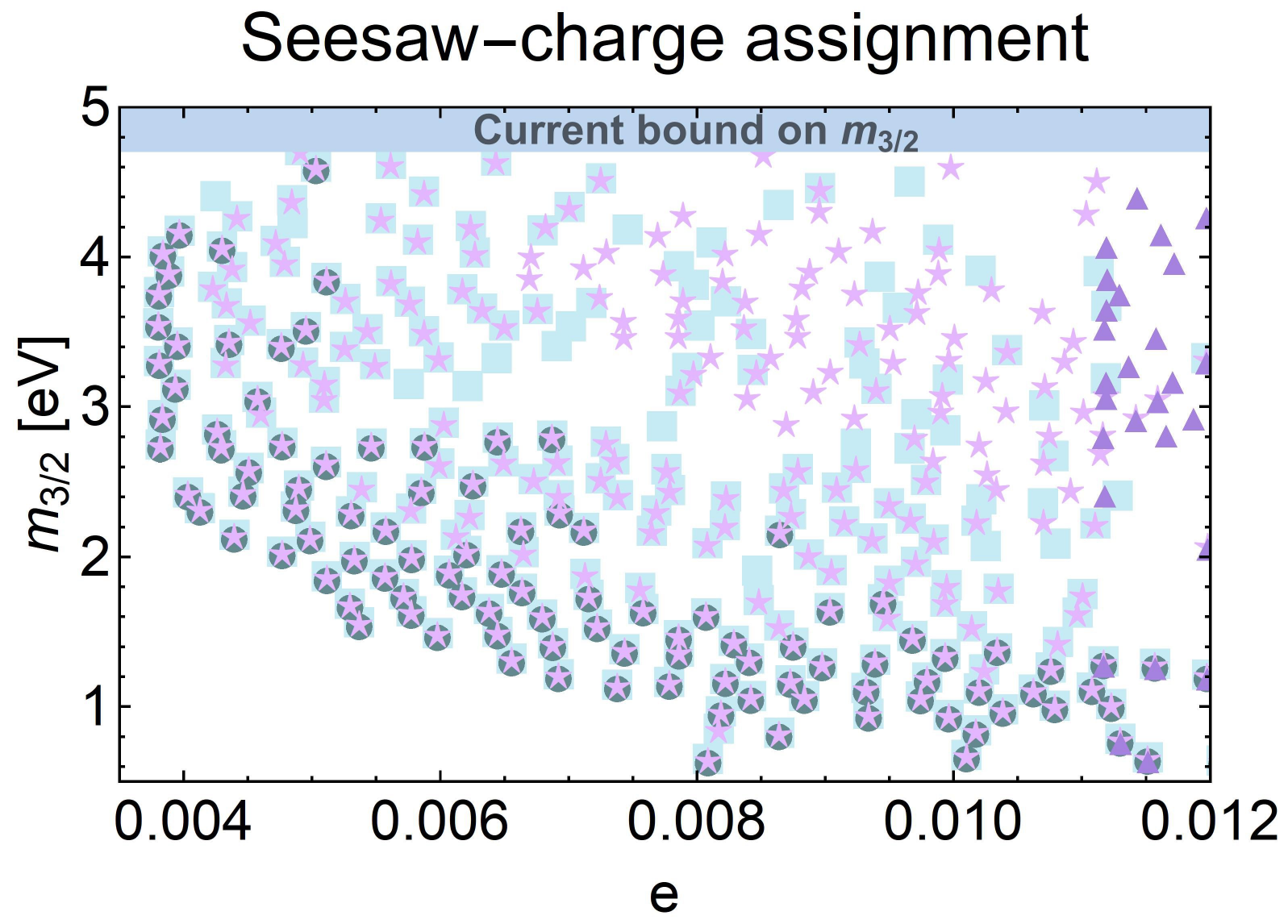}}
\centering
\subfloat{\includegraphics[width=0.44\textwidth]{SSQBadfig.pdf}}%
\hfil
\subfloat{\includegraphics[width=0.44\textwidth]{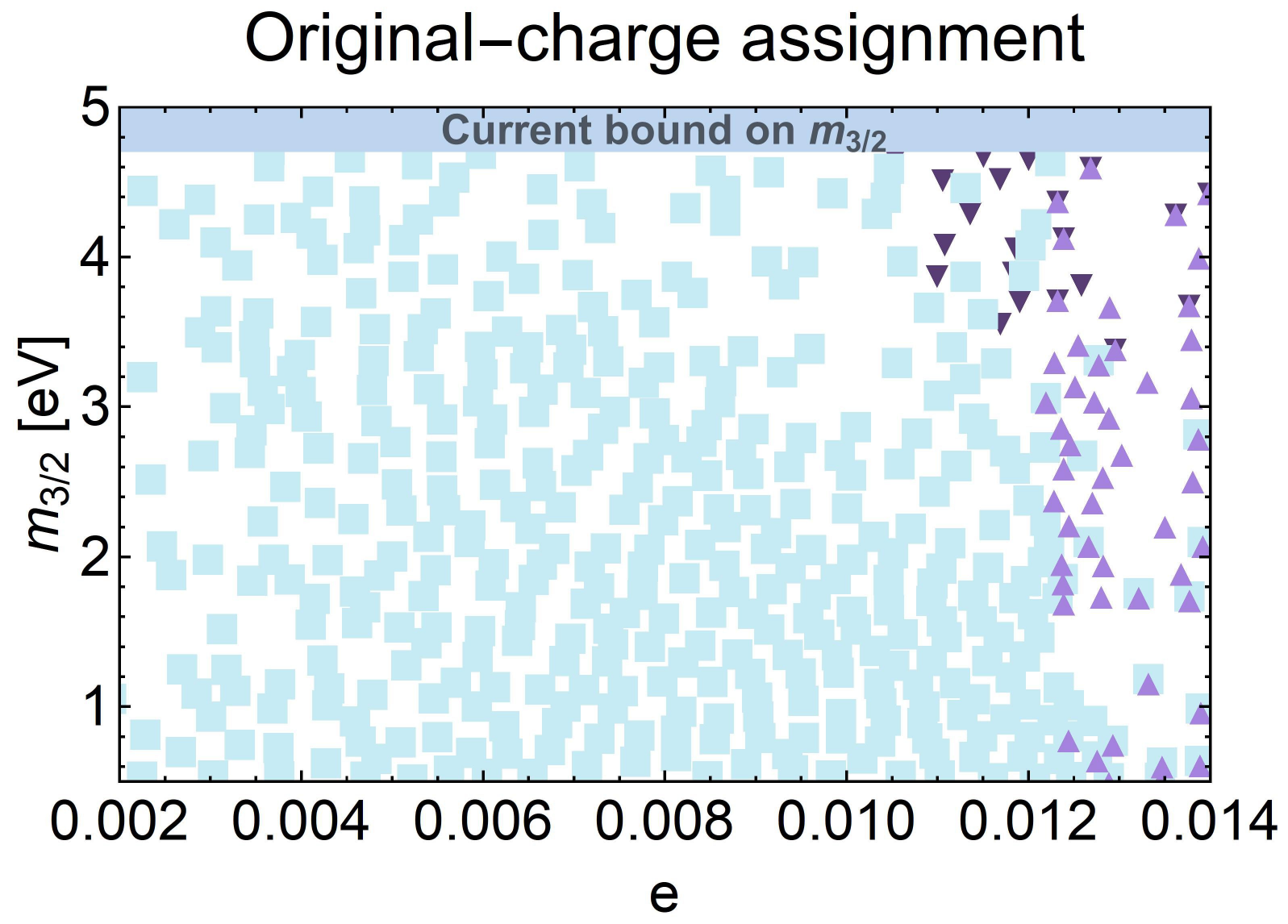}}%
\hfil
\subfloat{\tikz\node[minimum height=\ttemph]{\includegraphics[width=0.1\textwidth]{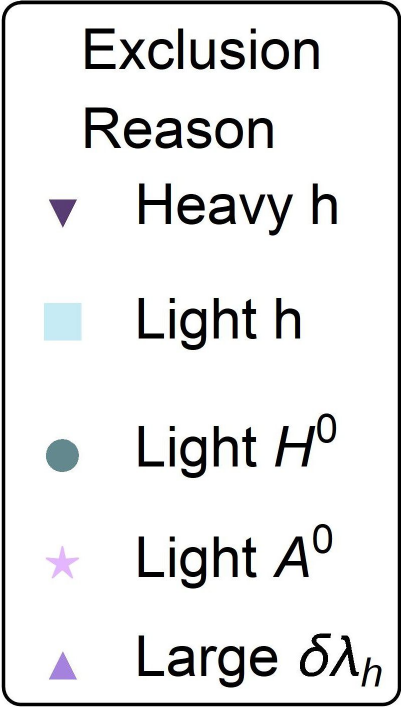}}; }%
\caption{$m_{3/2}$ vs.~$e$ for models with the seesaw-charge and original assignments
which are excluded by the aforementioned constraints. Excluded points
are shown along with their reason for exclusion when the Higgs is
calculated at the 3-loop level in SOFTSUSY.}
\label{fig:badfig}
\end{figure*}

Fig.~\ref{fig:badfig} shows why
representative points in parameter space are excluded by the aforementioned
constraints for the seesaw-charge and original-charge assignments
when we calculate the Higgs mass at the 3-loop level. The most exclusive
constraint is getting the Higgs mass correct. Points toward the bottom
and left of these figures correspond to a Higgs which is too light.
For the seesaw-charge assignment, the other MSSM Higgs scalars are
often too light when the Higgs is too light. Points to the far right
are excluded because $\delta\lambda_{h}$ from Eq.~\ref{eq:dellam}
is too large and the MSSM approximation breaks down.

\begin{figure}[!htp]%
\centering
\subfloat{\includegraphics[width=\columnwidth]{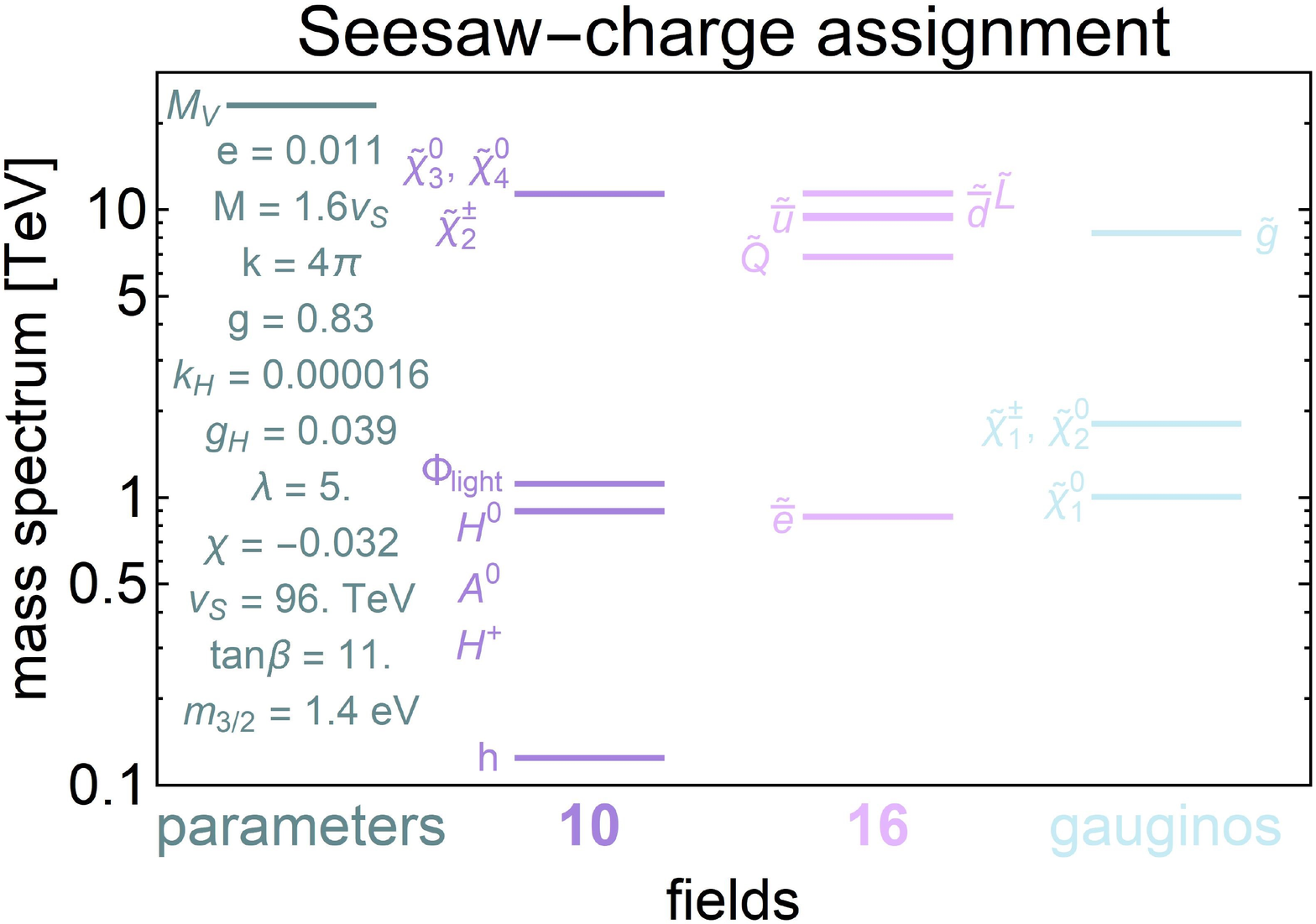}}%
\hfil
\subfloat{\includegraphics[width=\columnwidth]{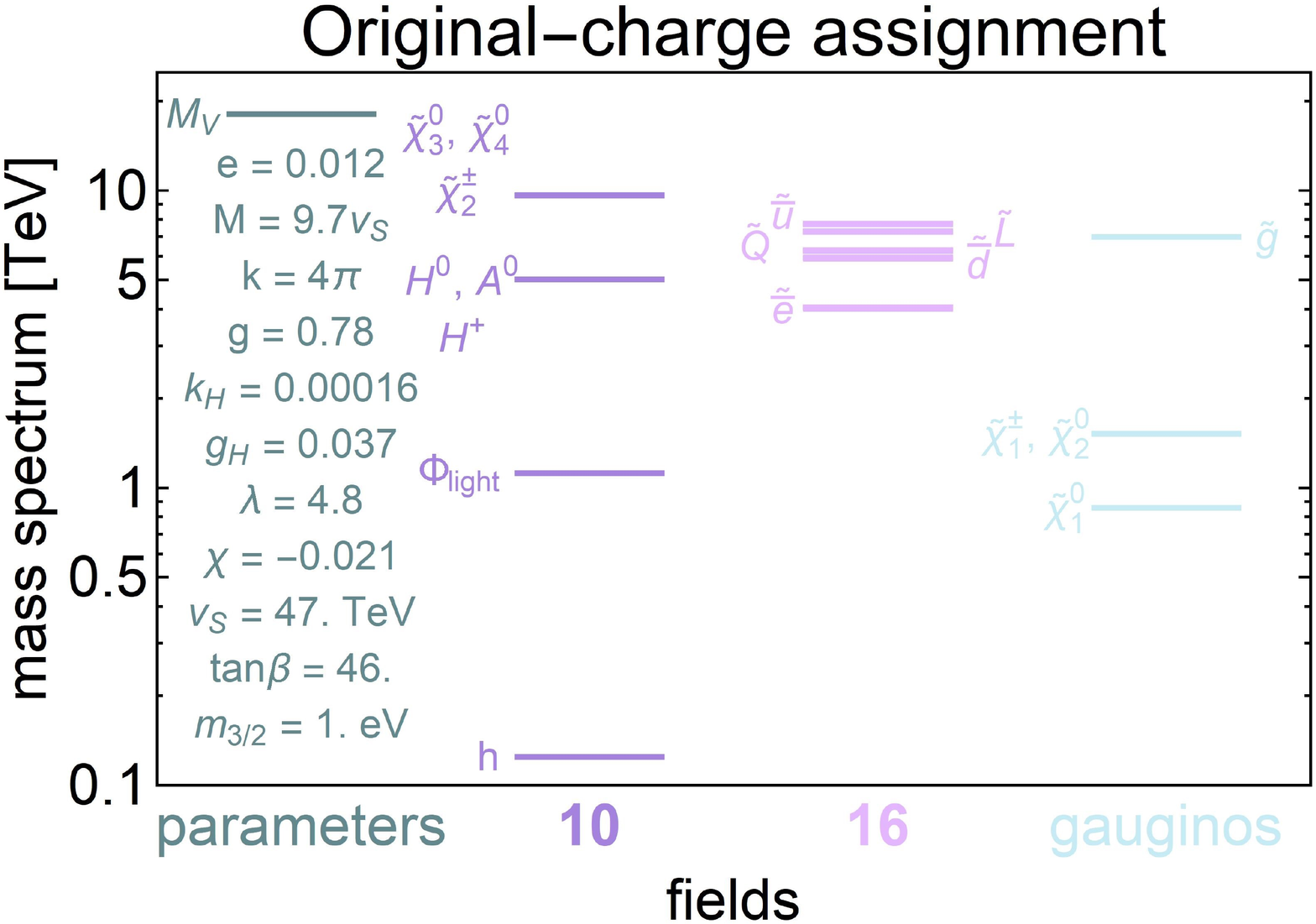}}%
\caption{Low-lying spectra for the seesaw-charge and original-charge models
with the lightest gravitino when $m_{h}$ is calculated at the 2-loop level.}
\label{fig:specfig}
\end{figure}

The viable models with the lightest gravitinos are illustrated in
Fig.~\ref{fig:specfig}. The equivalent
figures with $m_{h}$ calculated at the 3-loop level are similar, with
their spectra shifted slightly up. 

SOFTSUSY requires $\mu$ to be 9.74 TeV and 11.4 TeV for the original-
and seesaw-charge assignments, respectively, reducing the $\mu$ problem
from the Planck scale to the 10-TeV scale. One might be confused that
the gauginos are near the sfermions. This is simply an artifact of
maximizing the gaugino masses in units of $v_{S}$ before EWSB. The
lightest gravitino masses for these 3-loop calculations are $1.9$
eV and 2.3 eV for the original- and seesaw-charge assignments, respectively.
Optimistically, we see that we can obtain gravitino masses as light
as 1.0 eV and 1.4 eV for the original- and seesaw-charge assignments
when we calculate the Higgs at 2-loop level. Even being conservative
and calculating the Higgs mass at the 3-loop level still yields a
sufficiently light gravitino to evade the current cosmological bound
$m_{3/2}<4.7\,\text{eV}$.

\section{Conclusion - Discussions}

We expand on the previous work of Hook and Murayama by finding models
with lighter gravitinos, calculating EWSB, and enabling EWSB in models
with the new, seesaw-charge assignment by introducing $U(1)$ kinetic
mixing. Unlike typical, gauge-mediated models, we find that our vector-mediated
models are cosmologically viable with light gravitinos. As can be
extrapolated from Fig.~\ref{fig:goodfig}, going to larger
values of $e$ generally allows for lighter gravitinos. Since we use
SOFTSUSY to approximate EWSB in our models, we only consider $e\apprle0.01$
to avoid invalidating this approximation. If we calculate EWSB without
SOFTSUSY, we cannot consider values of $e$ much larger because the
vacuum and thermal tunneling rates begin to matter. For values of
$e\apprge0.1$, we can no longer choose $k$ and $g$ to maximize
the gaugino masses with respect to $v_{S}$. Satisfying the gaugino-mass
bounds then increases $v_{S}$ which increases the gravitino mass.
It is very interesting that the best we can do is not too far away
from current cosmological limits. Near-future improvements in cosmological
data, such as improvements in cosmic shear measurements at DES and
Hyper Suprime-Cam, could completely rule out low-energy SUSY breaking. 
\begin{acknowledgments}
R. M. thanks Simon Knapen for many useful discussions and an introduction
to SOFTSUSY. R. M. thanks Jason Evans and Marcin Badziak for helpful discussions.
R. M. also thanks Katelin Schutz for comments on drafts. H. M. and
R. M. thank T. T. Yanagida for discussions. R. M. is supported by
an NSF graduate research fellowship. A. H. is supported by NSF Grant
PHYS-1316699 and DOE Grant DE- SC0012012. This material is based upon
work supported by the National Science Foundation Graduate Research
Fellowship Program under Grant No. DGE 1106400. HM was supported by
the U.S. DOE under Contract DE-AC02-05CH11231, and by the NSF under
grants PHY-1316783 and PHY-1638509. HM was also supported by the JSPS
Grant-in-Aid for Scientific Research (C) (No.~26400241 and 17K05409),
MEXT Grant-in-Aid for Scientific Research on Innovative Areas (No.
15H05887, 15K21733), and by WPI, MEXT, Japan. Any opinions, findings,
and conclusions or recommendations expressed in this material are
those of the author(s) and do not necessarily reflect the views of
the National Science Foundation. 
\end{acknowledgments}

 \bibliographystyle{kp}

\begingroup\raggedright\endgroup

\end{document}